\documentclass[aps,twocolumn]{revtex4-2}
\usepackage[colorlinks]{hyperref}
\usepackage[utf8]{inputenc}
\usepackage{mathtools}

\begin{document}

\title{Microscopic composite systems bound by strong gravity in extra dimensions as candidates for dark matter}
\author{V. V. Flambaum}
\affiliation{School of Physics, University of New South Wales, Sydney 2052, Australia}

\begin{abstract}

In the Arkani-Hamed-Dimopoulos-Dvali (ADD) model with $n$ extra compactified dimensions, the gravitational potential scales as $1/r^{n+1}$ and becomes significantly stronger at short distances. We investigate the possibility of forming small-sized composite systems of Standard Model particles bound by this potential. Such bound states, composed of quarks, neutrinos, axions, or other particles, exhibit a small cross-section-to-mass ratio, making them viable candidates for dark matter.
\end{abstract}
	
\maketitle

\section{Introduction}

Observations indicate that  our universe has three non-compact spatial dimensions, with all forces and particles operating inside these dimensions. However, there are popular theoretical models with extra spatial dimensions such as string theories - see e.g. \cite{Witten1995}. These models have been motivated by search for theory unifying all interactions and producing finite results which do not require hiding infinities using the renormalisation procedures.     

In this paper, we examine the Arkani-Hamed-Dimopoulos-Dvali (ADD) model \cite{ADD1998,AADD1998,A1990}  which aimed at solving the hierarchy problem ( in this case it is a huge  difference  between the strength of gravity and strengths of other interactions between  elementary particles) by proposing that gravity can propagate through $n$ extra spatial compact dimensions as well as the regular three non-compact dimensions. In ADD model the observed Newton gravitational  law in our three dimensions will be significantly  strengthened  at distance smaller than the size of extra dimensions $R$ 
(see also Randall-Sundrum models \cite{RS_Model1_1999,RS_Model2_1999}, where the parameter is the distance between the "branes", parallel universes embedded in the multi-dimensional bulk).

The Newtonian gravitational potential will change from $1/r$ to a more singular $1/r^{n+1}$ dependence at distances smaller than the size of the extra dimensions, $R$. This may be easily explained by the Gauss integrated  flux formula for the gravitational force since size of the "surface" in $(3+n)D$ is proportional to $r^{2+n}$, therefore to have a constant flux the radial field strength $\propto 1/r^{2+n}$    \cite{ADD1998} (see also the  Randall-Sundrum multi-dimensional models \cite{RS_Model1_1999,RS_Model2_1999} which have similar small-distance gravity). Search for macroscopic effects has not found any deviation from Newton  law. Gravity force has been observed to obey the inverse-square law down to the $\mu$m scale \cite{TanPRL2020,PhysRevLett.124.101101}, see also  Refs.~\cite{Kapner2007,Adelberger2007,Chen2016,Vasilakis2009,Terrano2015}. Smaller distances may be probed in atomic phenomena.  Several papers \cite{FengCPL2006,FengIJTP2007,DahiaPRD2016,ZhiGang2008,WangMPLA2013,MunroLaylim} have attempted to calculate this effect of extra dimensions on spectra of simple atomic systems and placed constraints on $R$ through first order perturbation theory. The problem is that the results  strongly depend on an  unknown (and practically arbitrary) cut-off parameter which has to be introduced to obtain finite energy shift produced by  the gravitational potential $g/r^{n+1}$ for $n>1$. Another problem is that perturbation theory is not applicable if the cut-off parameter is too small and the gravitational potential near the cut-off exceeds the Coulomb potential.  These problems have been avoided  in Ref. \cite{Salumbides2015} where the authors  
studied effects of potential $g/r^{n+1}$ between nuclei in H$_2$, D$_2$ and HD$^+$ molecules. Since nuclei in molecules are separated by distance exceeding Bohr radius $a_B$, there is no need in cut-off parameter here.  However, the  effects of the  potential $g/r^{n+1}$  are much bigger for subatomic distances which give the main contribution to the energy shift in atoms. 

In Ref. \cite{MunroLaylim} the cut-off problem was solved empirically,  using  simultaneously  two experimental facts: absence of the collapsed gravitational bound states of two elementary particles (two electrons, electron and quark or two quarks) and difference between the measured and calculated (using QED) transition energy in simple atomic systems (muonium, positronium, hydrogen and deuterium). Indeed, the highly singular nature of the ADD gravitational potential $g/r^{n+1}$ introduces a fall-to-centre problem, where there must exist some new physics mechanism that cuts off the gravitational potential below some cut-off radius $r_c$. In this way Ref. \cite{MunroLaylim} obtained  conservative  limits on the size of extra dimensions by excluding the area of unknown physics at distances smaller than the electroweak scale.

In the present paper we want to study a possibility of small-sized bound states of a large number of  elementary particles. Such states are not excluded by experiment. Moreover, such states may be attractive candidates for dark matter. These states may remain undetected if they possess a sufficiently small cross section relative to their mass.
The most notable example of such a model is strangelets, hypothetical droplets of quark matter stabilized by strange quarks \cite{Witten84,Farhi,Glashow}. More generally, macroscopic composite dark matter particles were studied in a series of works \cite{Starkman1,Starkman2,Starkman3,Starkman4,Starkman5,Starkman6,Starkman7,Starkman8,Starkman9,Starkman10,Caloni,FS0,FS1,FS2,Garry,Garry2025,Budker2020,Budker2022}.
An extension of the strangelet model developed by A. Zhitnitsky and his collaborators in a series of papers \cite{DW1,DW2,DW3,Zhitnitsky_2003,CCO}, dubbed axion-quark nuggets (AQNs). Unlike strangelets, this model assumes an axion-pion domain wall that creates pressure on the quark core, providing stability against decay into baryonic matter. Another notable aspect of this model is the assumption that, in addition to quark nuggets (QNs), there are antiquark nuggets that comprise all antimatter, making the total number of quarks and antiquarks in the Universe equal \cite{Zhitnitsky_2003,CCO,Basymmetry,WMAPhaze}.
This idea offers a possible solution to the problem of matter-antimatter asymmetry in the Universe.

However,  there is a problem of stability of such states since their energy must be smaller than the energy of nuclear matter. In the present paper we investigate a possibility that this stability is provided by the singular gravitational potential  $g/r^{n+1}$ for $n>0$.

\section{Overview of The ADD Model}
The ADD model was first proposed in Ref. \cite{ADD1998} and introduced $n$ extra spatial compactified dimensions of size $R_{(n)}$. There is also a change to the Planck mass, where a higher dimensional Planck mass $M$ is defined with respect to $R$ and replaces the observed three-dimensional Planck mass $M_{Pl}$
\begin{equation}
    \left(\frac{c M_{Pl}}{\hbar}\right)^2 = R_{(n)}^n \left(\frac{c M}{\hbar}\right)^{n+2}.
\end{equation}

This means that Gauss's Law can be applied to the gravitational potential energy for two masses $m_1$ and $m_2$ separated by distance  $r$:
\begin{eqnarray}
\label{e:V_ADD}
    V(r) =\left\{ 
    \begin{array}{ll}
         -\frac{G m_1 m_2}{r} & \mathrm{ for }\,\, r\gg R_{(n)},\\
         -\frac{G m_1 m_2 R_{(n)}^n}{r^{n+1}} & \mathrm{ for }\,\, r\ll R_{(n)},
  \end{array}
  \right.
\end{eqnarray}
where $G=6.674\times10^{-11}$ m$^3$ kg$^{-1}$ s$^{-2}$ is the regular gravitational constant. The long-range component of the potential  matches the regular Newtonian gravitational potential that is known to be accurate from $\mu$m scales to astronomical scales.  Small distance gravity tests   \cite{TanPRL2020,PhysRevLett.124.101101} indicate that $R_{(n)}>$30 $\mu$m. However, the short-range component has a more singular form, therefore this is the region where gravitational potential can become significantly stronger with extra dimensions.

\section{Gravitationally bound  fermions in the ground state}

Let us consider an electrically  neutral system of $N$ fermionic particles bound by modified gravitational interaction (\ref{e:V_ADD}). In the case of non-modified gravity, such system was studied in Ref.~\cite{Ruffini} where a numerical solution for the Schr\"odinger-Poisson equation was found. Here, we will rather follow a variational approach by choosing a specific ansatz for the particle density which then should minimize the total energy of the system meaning that such an ansatz should describe the ground state. Following Ref.~\cite{BoseStar}, we consider a spherically symmetric distribution with a Gaussian profile for the particle number density 
\begin{equation}
    n(r) = N \left( \frac{2}{\pi R^2} \right)^{3/2}
    e^{-2r^2/R^2}\,,
    \label{psi-Gaussian}
\end{equation}
where $R$ is the parameter to be found from the condition of minimum of energy. This density is normalized to the number of particles $N$,
\begin{equation}
    \int d^3r \,n(r) = 4\pi\int_0^\infty dr\,r^2n(r) =N\,.
\end{equation}
The root-mean-squared radius calculated with the density  (\ref{psi-Gaussian}) 
\begin{equation}
    r_\text{rms}  \equiv \left[4\pi \int_0^\infty dr\,r^4n(r)\right]^{1/2} =\frac{\sqrt3 R}{2}
    \label{r-rms-Gaussian}
\end{equation}
shows that $R$ may be considered as the characteristic size of the solution.

In systems of a very small radius $R$,  particle momentum $p \gtrsim 1/R$ exceeds the particle rest mass, and a relativistic treatment is required. In the ideal ultrarelativistic Fermi gas, the kinetic energy is expressed in terms of particle number density $n$ as follows  \cite{Landau5}:
\begin{equation}\label{kinErel}
E_\mathrm{kin} = \frac34 (3\pi^2)^{1/3} \int d^3r\,n^{4/3}\,. 
\end{equation}
Integrating the Gaussian distribution of density (\ref{psi-Gaussian}) we obtain 
\begin{equation}
\label{kinErelR}
E_\mathrm{kin} = \frac{9\times 3^{5/6}\pi^{1/6}N^{4/3}}{16\sqrt{2}R}\approx 1.2\frac{N^{4/3}}{R}\,.
\end{equation}

Direct contribution to the gravitational energy is 
\begin{equation}
\label{GravErel0}
E_\mathrm{grav}= \frac12\int n(r_1) n(r_2)V(\mathbf{r}_1-\mathbf{r}_2) d^3r_1 d^3r_2\,,
\end{equation}
where $V(r)$ is the modified gravity potential (\ref{e:V_ADD}).

Equation (\ref{GravErel0}) may be simplified by making change  of the variables ${\bf r'}=({\bf r_1}+{\bf r_2})/2$, ${\bf r}=({\bf r_2}-{\bf r_1})$ and integrating over $r'$ using the explicit expression for the particle density (\ref{psi-Gaussian}),
\begin{equation}
\label{GravEr}
E_\mathrm{grav}= \frac{N}{2^{5/2}}\int d^3r \,n(r/\sqrt{2})V(r)\,.
\end{equation}
The calculations of this integral for the potential (\ref{e:V_ADD}) is specific for different values of the number of extra dimensions $n$. Consider these cases separately.

The case $n=1$ is special because the corresponding integral of (\ref{e:V_ADD}) with the density (\ref{psi-Gaussian}) is convergent,
\begin{equation}\label{grav1}
    E_\mathrm{grav}\simeq -\frac{N^2Gm^2 R_{(1)}}{R^2}\,.\qquad (n=1)
\end{equation}
Here we extended the short-distance gravitational potential $V(r)= - Gm^2 R_{(1)}/r^2$ to the entire space, although for $r>R_{(1)}$ the long-distance gravitational potential $V(r) = -Gm^2/r$ could be used. It is possible to show that this systematic underestimation of the energy is minor and may be ignored for the for sake of simplicity of the analytic result.

In the case $n>1$ gravitation energy integral is divergent at small distance. We regularize this integral by introducing a small-$r$ cut-off parameter $r_c\ll \mathrm{min}(R,R_{(n)})$, and notice that the leading contributions to the energy are given by the singular terms at small $r_c$,
\begin{subequations}
\label{gravAll}
\begin{align} \label{grav2}
E_\mathrm{grav} &\simeq -\frac{2 N^2 Gm^2R^2_{(2)}}{\sqrt{\pi}R^3} \ln{\frac{R}{r_c}}, && (n=2)\\
E_\mathrm{grav} &\simeq -\frac{4 N^2 Gm^2R^n_{(n)}}{2^{n/2}\sqrt{\pi}(n-2)r_c^{n-2}R^3}  , && (n>2)\,.
\label{gravn}
\end{align}
\end{subequations}

In all $n$ cases the minimum of total energy $E=E_\mathrm{kin}+E_\mathrm{grav}$ is achieved for $R \sim  r_c$. Here, we assume that in small-sized  neutral (and color-neutral) systems, the modified gravitational interaction given by Eqs.~(\ref{grav1}) and (\ref{gravAll}) exceeds the contributions of the strong and electroweak interactions. There may be two potential reasons for this gravity dominance. First, electrostatic potential is screened in neutral systems, while gravitational interaction cannot be screened due to its universal attractive nature  (compare with ordinary gravitationally bound celestial bodies). 
Second, the modified gravitational interaction has a stronger singularity at short distances, enhancing its contribution.

Indirect effects of other interactions can still be accounted for through an effective mass in the gravitational energy formulas. For instance, the constituent quark mass ($m_q \approx m_p/3 \approx 300$ MeV) includes the energy of the gluon field inside the proton. This gluon energy effectively contributes to the gravitational energy in ordinary gravitationally bound celestial systems.

The condition for bound state $E_\mathrm{grav}+ E_\mathrm{kin}<0$ gives us an estimate for the minimal number of fermions which may form a gravitational bound state of small size  $R \sim  r_c$:  
\begin{equation}\label{NcFermiRel}
N^{2/3} \gtrsim \frac{r_c^{n}}{(R_{(n)})^n }\frac{1}{Gm^2}\,.  
\end{equation}
A similar estimate for the ground state of  bosons and non-identical fermions is
\begin{equation}\label{NcBoseRel}
N \gtrsim \frac{r_c^{n}}{(R_{(n)})^n }\frac{1}{Gm^2}\,.  
\end{equation}
These equations contain two unknown parameters, $R_{(n)}$ and $r_c$.  
To estimate the ratio $\frac{r_c^{n}}{(R_{(n)})^n }$ we may use the following argument. The idea of this approach is an approximate equality of different interactions at distance $r_c$. Therefore, we may assume that the gravitational interaction and electromagnetic, weak and strong interactions  at the distance $r_c$ are comparable (within one order of magnitude):  
\begin{equation}
\label{gravElectr}
    \frac{G m^2 R_{(n)}^n}{r_c^{n+1}} \sim \frac{\alpha}{r_c}\,,
\end{equation}
where $\alpha\approx1/137$ is the fine structure constant. This gives us an estimate for the number of identical fermion which are able to  form small size gravitational bound state:
\begin{equation}\label{minN}
N \gtrsim \frac{1}{\alpha^{3/2}}\approx1600\,.
\end{equation} 


The naive estimates for the minimal value of $N$ presented above serve as an illustration of a possible outcome. A proper theory should incorporate the generalized Einstein equations, which account for extra compactified dimensions, as well as a new short-distance (high-energy) framework that defines the cut-off parameter $r_c$. A detailed discussion of such theories is beyond the scope of this work.

It is also worth noting that a certain degree of self-interaction among dark matter particles is considered desirable to explain some astrophysical observations. For further discussion, see, for example, Ref.~\cite{DMselfinteraction} and references therein.

\section{Possible mechanisms for the  formation and survival  of the quark nuggets} 

Witten \cite{Witten84} proposed that neutral or near-neutral quark nuggets could form during a first-order QCD (quark-hadron) phase transition at a critical temperature of  $T_c \sim$ 150 MeV.  In the final stage of such a transition, shrinking regions of the quark phase would collapse, compressing the quark matter inside.   The singular gravitational potential considered in this work would help stabilize these compressed regions. A challenge to this scenario, however, is that current QCD results indicate the QCD transition is not first-order, which may inhibit nugget formation through this route.

A. Zhitnitsky and  collaborators \cite{DW1,DW2,DW3,Zhitnitsky_2003,CCO}  proposed an alternative mechanism in which collapsing axion domain walls sweep up quarks and antiquarks into nuggets and antinuggets. This scenario does not rely on the QCD quark-hadron transition being first order.

Another possibility is that quark nuggets form through a mechanism analogous to primordial black hole production from early-universe density fluctuations (see, e.g., \cite{BlackHole}). If a fluctuation is insufficiently overdense to collapse into a black hole, it could instead compress into a stable quark nugget.

  To prevent collapse into a black hole, the system radius must satisfy $R \sim r_c >r_g$, where  $r_g=2 G M  \sim N G m$ is the gravitational  (Schwarzschild) radius. For example, for the constituent quark mass  $m= 300$ MeV, we obtain
 $N < 10^{38} r_c/$fm.  If  $N$ exceeds this limit,  so that $r_g >r_c$,  the object would instead collapse to a black hole.
 
We should also address a problem of the quark nuggets evaporation in the hot primordial plazma.  Following Ref.~\cite{neutronevaporation}, the neutron evaporation rate from a quark-nugget surface of area $4 \pi R^2$ is\begin{equation}
\frac{dB}{dt}=-\frac{2}{\pi} m_n T^2 R^2\,e^{-I_n/T},
\end{equation} 
where $B=3N$ is the barion charge of the quark nugget, $m_n$ is the neutron mass, $T$ is the temperature, $R$ is the nugget radius, and $I_n$ is the neutron binding energy.  The derivation of this expression follows the logic used for the black-body radiation law: in thermal equilibrium, the rate of particle emission equals the rate of absorption (see, e.g., Ref.\cite{Landau5}). Ref.\cite{MadsenHeiselberg} extended the analysis to include the emission of other hadrons. Integrating the evaporation rate from the moment of nugget formation at $T \sim 100$ MeV to $T=0$  yields a survival bound for Witten-type nuggets $B> 10^{46}$   \cite{MadsenHeiselberg}. Allowing for the reabsorption of ambient hadrons can lower this bound considerably.  It was stated in Ref.  \cite{MadsenHeiselberg} that at temperature $T< 15$ MeV reabsorption dominates over evaporation.  Refs.~\cite{Madsen1998,Zhitnitsky_2003} argue that the depletion of uu and dd quarks near the surface (caused by neutron evaporation), together with colour-superconducting effects and related mechanisms, can relax the survival limit to $B> 10^{30}$.  
For antinuggets, an additional constraint of $|B|> 10^{25}$ arises from the annihilation of incident nucleons on the nugget surface \cite{Zhitnitsky2019,FS2}.  Ref.~\cite{Zhitnitsky2019} further argues that quark nuggets formed at a temperature  $T \sim 40$ MeV; under such conditions, annihilation dominates over evaporation, making the latter process comparatively unimportant.

Note that in our case of the gravitationally bound nuggets, the surface area  $ A_G= 4 \pi R^2  \sim 4 \pi r_c^2 $ is extremely  small compared to the Witten-Zhitnitsky  nugget area  $ A_W=4 \pi  B^{2/3} $ fm$^2$. This drastic difference strongly suppresses the evaporation rate. For example, choosing $r_c  =1/$TeV (the largest reasonable value) gives a suppression factor $A_G/A_W \sim 10^{-8}/B^{2/3}$,  yielding  $10^{-32}$ for $B=10^{36}$ and $10^{-20}$ for $B=10^{18}$ .

Note also that the emission of particles whose wavelengths exceed the nugget size is suppressed relative to standard black-body radiation; see Ref.~\cite{FS1} for the nugget thermal emission formula that applies to photons of arbitrary energy. This suppression is especially important for gravitationally bound nuggets, whose radii are far smaller than the typical particle wavelengths at the relevant temperatures.

In summary, gravitationally bound nuggets are more likely to survive the hot primordial plasma than Witten-Zhitnitsky nuggets.
  
\section{Limits from existing observations and detection prospects}

The composite objects considered here - whether made of quarks, neutrinos, axions, or other elementary particles - possess a very small cross-section-to-mass ratio, making them plausible dark-matter candidates. For a quark nugget of extremely small radius, $\sim r_c \ll$ fm, the interaction cross section is set by the strong interaction radius, $\sigma \sim \text{fm}^2$.  Assuming a quark nugget mass $M \sim B m_p $, this cross section satisfies the astrophysical limit on the ratio $\sigma/M < 0.1$ cm$^2/$g $\sim 10\,  \text{fm}^2/\text{GeV}$\cite{PDG} for any barion number $B$.

While laboratory limits on quark nugget parameters can be tighter, they are effectively irrelevant for dark-matter candidates with very large masses, because the chance that such massive particles traverse a detector is exceedingly small. The number density of heavy nuggets scales as $\rho/M$, where $\rho \sim 0.4\, \text{GeV}/\text{cm}^3$ is the local dark matter mass density near the Solar System.
 For a nugget mass $M \sim B m_p $ one would expect roughly one nugget per year to cross a  m$^2$  detector if $B \sim 10^{19}$ and nuggets make up all of the dark matter.


Current   direct-detection constraints  on WIMPs (weakly interacting massive particles)   extend only up to a WIMP mass of about  1 TeV \cite{PDG}.  However, the gravitational quark nugget has  the strong interaction cross section $\sim$ fm$^2$ which exceeds expected  WIMP cross section by many orders of magnitude.  Consequently, existing dark-matter detectors
 - such as  LZ, XENONnT, PandaX-4T, DEAP-3600, DarkSide-50, CRESST-III, PICO-40L,  COSINE-100, ANAIS-112 
 -  could in principle be sensitive to gravitational quark nuggets with baryon numbers up to  $B \sim  10^{18}$.  

To probe even larger $B$ values, one can turn to very large detectors such as IceCube. The absence of signals there already implies $B>10^{25}$  for Witten-Zhitnitsky nuggets \cite{Zhitnitsky2019,FS2} (see also the review in Ref.~\cite{ZhitnitskyReview}). IceCube, as well as other kilometer-scale experiments like KM3NeT-ARCA/ORCA and Baikal-GVD, reaches a sensitivity floor set by weak interactions; with their enormous volumes, these detectors should easily be sensitive to gravitational nuggets, whose strong-interaction cross sections are much larger.


 Moreover, the entire Earth can effectively serve as a detector for extremely heavy quark nuggets with $B>10^{25}$. Such nuggets would deposit substantial energy while traversing the planet, producing observable signals. They could be detected via meteor-search radars that register the ionisation trails left by the nuggets \cite{Radar,Garry}, and  via
 seismological networks distributed across all continents, the oceans, and even the Moon, which would record the seismic waves generated by their passage \cite{Herrin1,Herrin2,Herrin3,Starkman9,Budker2020,Budker2022,Garry2025}. Recent work \cite{Garry2025} reviews the current constraints on Witten-Zhitnitsky nuggets and antinuggets and provides an extensive list of references. The limits are particularly stringent for antinuggets because matter-antimatter annihilation would release large amounts of energy and produce an accompanying flux of neutrinos.
 

However, these bounds do not translate directly to gravitationally bound quark nuggets, whose interaction cross section is smaller by a factor of $B^{2/3}$.  Consequently, the energy released by a passing gravitational nugget is suppressed by the same factor relative to a Witten-Zhitnitsky nugget. The existing sensitivity of seismic, radar, and neutrino searches must therefore be re-evaluated. Our preliminary estimates indicate that the current measurements discussed in Ref.~\cite{Garry2025} (and the references therein) do not impose any additional, significant limits on gravitational quark nuggets or antinuggets.

As discussed above, developing a comprehensive theory for gravitational nuggets - including their parameters, evaporation, radiation, and interaction with matter - will involve many facets and is left for future work.



\vspace{2mm}
\textit{Acknowledgements.}--- 
I am grateful to Igor Samsonov for his contributions during the initial stages of this work. I thank Jasmine Azizi, Harry Mangos, Matilda Stewart and Iris Wu for their interest in this work. 
 The work was supported by the Australian Research Council Grant No.\ DP230101058.

%


\begin{thebibliography}{49}%
\makeatletter
\providecommand \@ifxundefined [1]{%
 \@ifx{#1\undefined}
}%
\providecommand \@ifnum [1]{%
 \ifnum #1\expandafter \@firstoftwo
 \else \expandafter \@secondoftwo
 \fi
}%
\providecommand \@ifx [1]{%
 \ifx #1\expandafter \@firstoftwo
 \else \expandafter \@secondoftwo
 \fi
}%
\providecommand \natexlab [1]{#1}%
\providecommand \enquote  [1]{``#1''}%
\providecommand \bibnamefont  [1]{#1}%
\providecommand \bibfnamefont [1]{#1}%
\providecommand \citenamefont [1]{#1}%
\providecommand \href@noop [0]{\@secondoftwo}%
\providecommand \href [0]{\begingroup \@sanitize@url \@href}%
\providecommand \@href[1]{\@@startlink{#1}\@@href}%
\providecommand \@@href[1]{\endgroup#1\@@endlink}%
\providecommand \@sanitize@url [0]{\catcode `\\12\catcode `\$12\cat code
  `\&12\catcode `\#12\catcode `\^12\catcode `\_12\catcode `\%12\relax}%
\providecommand \@@startlink[1]{}%
\providecommand \@@endlink[0]{}%
\providecommand \url  [0]{\begingroup\@sanitize@url \@url }%
\providecommand \@url [1]{\endgroup\@href {#1}{\urlprefix }}%
\providecommand \urlprefix  [0]{URL }%
\providecommand \Eprint [0]{\href }%
\providecommand \doibase [0]{https://doi.org/}%
\providecommand \selectlanguage [0]{\@gobble}%
\providecommand \bibinfo  [0]{\@secondoftwo}%
\providecommand \bibfield  [0]{\@secondoftwo}%
\providecommand \translation [1]{[#1]}%
\providecommand \BibitemOpen [0]{}%
\providecommand \bibitemStop [0]{}%
\providecommand \bibitemNoStop [0]{.\EOS\space}%
\providecommand \EOS [0]{\spacefactor3000\relax}%
\providecommand \BibitemShut  [1]{\csname bibitem#1\endcsname}%
\let\auto@bib@innerbib\@empty
\bibitem [{\citenamefont {Witten}(1995)}]{Witten1995}%
  \BibitemOpen
  \bibfield  {author} {\bibinfo {author} {\bibfnamefont {E.}~\bibnamefont
  {Witten}},\ }\href {https://doi.org/10.1016/0550-3213(95)00158-o} {\bibfield
  {journal} {\bibinfo  {journal} {Nuclear Physics B}\ }\textbf {\bibinfo
  {volume} {443}},\ \bibinfo {pages} {85} (\bibinfo {year} {1995})}\BibitemShut
  {NoStop}%
\bibitem [{\citenamefont {Arkani–Hamed}\ \emph {et~al.}(1998)\citenamefont
  {Arkani–Hamed}, \citenamefont {Dimopoulos},\ and\ \citenamefont
  {Dvali}}]{ADD1998}%
  \BibitemOpen
  \bibfield  {author} {\bibinfo {author} {\bibfnamefont {N.}~\bibnamefont
  {Arkani–Hamed}}, \bibinfo {author} {\bibfnamefont {S.}~\bibnamefont
  {Dimopoulos}},\ and\ \bibinfo {author} {\bibfnamefont {G.}~\bibnamefont
  {Dvali}},\ }\href {https://doi.org/10.1016/s0370-2693(98)00466-3} {\bibfield
  {journal} {\bibinfo  {journal} {Physics Letters B}\ }\textbf {\bibinfo
  {volume} {429}},\ \bibinfo {pages} {263–272} (\bibinfo {year}
  {1998})}\BibitemShut {NoStop}%
  \bibitem{AADD1998} Ignatios Antoniadis, Nima Arkani-Hamed, Savas Dimopoulos, Gia Dvali.  Phys. Lett. B {\bf 436}, 257 (1998). 
  \bibitem{A1990} Ignatios Antoniadis, Phys. Lett. B {\bf 246}, 377 (1990). 
\bibitem [{\citenamefont {Randall}\ and\ \citenamefont
  {Sundrum}(1999{\natexlab{a}})}]{RS_Model1_1999}%
  \BibitemOpen
  \bibfield  {author} {\bibinfo {author} {\bibfnamefont {L.}~\bibnamefont
  {Randall}}\ and\ \bibinfo {author} {\bibfnamefont {R.}~\bibnamefont
  {Sundrum}},\ }\href {https://doi.org/10.1103/PhysRevLett.83.3370} {\bibfield
  {journal} {\bibinfo  {journal} {Phys. Rev. Lett.}\ }\textbf {\bibinfo
  {volume} {83}},\ \bibinfo {pages} {3370} (\bibinfo {year}
  {1999}{\natexlab{a}})}\BibitemShut {NoStop}%
\bibitem [{\citenamefont {Randall}\ and\ \citenamefont
  {Sundrum}(1999{\natexlab{b}})}]{RS_Model2_1999}%
  \BibitemOpen
  \bibfield  {author} {\bibinfo {author} {\bibfnamefont {L.}~\bibnamefont
  {Randall}}\ and\ \bibinfo {author} {\bibfnamefont {R.}~\bibnamefont
  {Sundrum}},\ }\href {https://doi.org/10.1103/PhysRevLett.83.4690} {\bibfield
  {journal} {\bibinfo  {journal} {Phys. Rev. Lett.}\ }\textbf {\bibinfo
  {volume} {83}},\ \bibinfo {pages} {4690} (\bibinfo {year}
  {1999}{\natexlab{b}})}\BibitemShut {NoStop}%
\bibitem [{\citenamefont {Tan}\ \emph {et~al.}(2020)\citenamefont {Tan},
  \citenamefont {Du}, \citenamefont {Dong}, \citenamefont {Yang}, \citenamefont
  {Shao}, \citenamefont {Guan}, \citenamefont {Wang}, \citenamefont {Zhan},
  \citenamefont {Luo}, \citenamefont {Tu},\ and\ \citenamefont
  {Luo}}]{TanPRL2020}%
  \BibitemOpen
  \bibfield  {author} {\bibinfo {author} {\bibfnamefont {W.-H.}\ \bibnamefont
  {Tan}}, \bibinfo {author} {\bibfnamefont {A.-B.}\ \bibnamefont {Du}},
  \bibinfo {author} {\bibfnamefont {W.-C.}\ \bibnamefont {Dong}}, \bibinfo
  {author} {\bibfnamefont {S.-Q.}\ \bibnamefont {Yang}}, \bibinfo {author}
  {\bibfnamefont {C.-G.}\ \bibnamefont {Shao}}, \bibinfo {author}
  {\bibfnamefont {S.-G.}\ \bibnamefont {Guan}}, \bibinfo {author}
  {\bibfnamefont {Q.-L.}\ \bibnamefont {Wang}}, \bibinfo {author}
  {\bibfnamefont {B.-F.}\ \bibnamefont {Zhan}}, \bibinfo {author}
  {\bibfnamefont {P.-S.}\ \bibnamefont {Luo}}, \bibinfo {author} {\bibfnamefont
  {L.-C.}\ \bibnamefont {Tu}},\ and\ \bibinfo {author} {\bibfnamefont
  {J.}~\bibnamefont {Luo}},\ }\href
  {https://doi.org/10.1103/PhysRevLett.124.051301} {\bibfield  {journal}
  {\bibinfo  {journal} {Phys. Rev. Lett.}\ }\textbf {\bibinfo {volume} {124}},\
  \bibinfo {pages} {051301} (\bibinfo {year} {2020})}\BibitemShut {NoStop}%
\bibitem [{\citenamefont {Lee}\ \emph {et~al.}(2020)\citenamefont {Lee},
  \citenamefont {Adelberger}, \citenamefont {Cook}, \citenamefont {Fleischer},\
  and\ \citenamefont {Heckel}}]{PhysRevLett.124.101101}%
  \BibitemOpen
  \bibfield  {author} {\bibinfo {author} {\bibfnamefont {J.~G.}\ \bibnamefont
  {Lee}}, \bibinfo {author} {\bibfnamefont {E.~G.}\ \bibnamefont {Adelberger}},
  \bibinfo {author} {\bibfnamefont {T.~S.}\ \bibnamefont {Cook}}, \bibinfo
  {author} {\bibfnamefont {S.~M.}\ \bibnamefont {Fleischer}},\ and\ \bibinfo
  {author} {\bibfnamefont {B.~R.}\ \bibnamefont {Heckel}},\ }\href
  {https://doi.org/10.1103/PhysRevLett.124.101101} {\bibfield  {journal}
  {\bibinfo  {journal} {Phys. Rev. Lett.}\ }\textbf {\bibinfo {volume} {124}},\
  \bibinfo {pages} {101101} (\bibinfo {year} {2020})}\BibitemShut {NoStop}%
\bibitem [{\citenamefont {Kapner}\ \emph {et~al.}(2007)\citenamefont {Kapner},
  \citenamefont {Cook}, \citenamefont {Adelberger}, \citenamefont {Gundlach},
  \citenamefont {Heckel}, \citenamefont {Hoyle},\ and\ \citenamefont
  {Swanson}}]{Kapner2007}%
  \BibitemOpen
  \bibfield  {author} {\bibinfo {author} {\bibfnamefont {D.~J.}\ \bibnamefont
  {Kapner}}, \bibinfo {author} {\bibfnamefont {T.~S.}\ \bibnamefont {Cook}},
  \bibinfo {author} {\bibfnamefont {E.~G.}\ \bibnamefont {Adelberger}},
  \bibinfo {author} {\bibfnamefont {J.~H.}\ \bibnamefont {Gundlach}}, \bibinfo
  {author} {\bibfnamefont {B.~R.}\ \bibnamefont {Heckel}}, \bibinfo {author}
  {\bibfnamefont {C.~D.}\ \bibnamefont {Hoyle}},\ and\ \bibinfo {author}
  {\bibfnamefont {H.~E.}\ \bibnamefont {Swanson}},\ }\href@noop {} {\bibfield
  {journal} {\bibinfo  {journal} {Phys. Rev. Lett.}\ }\textbf {\bibinfo
  {volume} {98}},\ \bibinfo {pages} {021101} (\bibinfo {year}
  {2007})}\BibitemShut {NoStop}%
\bibitem [{\citenamefont {Adelberger}\ \emph {et~al.}(2007)\citenamefont
  {Adelberger}, \citenamefont {Heckel}, \citenamefont {Hoedl}, \citenamefont
  {Hoyle}, \citenamefont {Kapner},\ and\ \citenamefont
  {Upadhye}}]{Adelberger2007}%
  \BibitemOpen
  \bibfield  {author} {\bibinfo {author} {\bibfnamefont {E.~G.}\ \bibnamefont
  {Adelberger}}, \bibinfo {author} {\bibfnamefont {B.~R.}\ \bibnamefont
  {Heckel}}, \bibinfo {author} {\bibfnamefont {S.}~\bibnamefont {Hoedl}},
  \bibinfo {author} {\bibfnamefont {C.~D.}\ \bibnamefont {Hoyle}}, \bibinfo
  {author} {\bibfnamefont {D.~J.}\ \bibnamefont {Kapner}},\ and\ \bibinfo
  {author} {\bibfnamefont {A.}~\bibnamefont {Upadhye}},\ }\href@noop {}
  {\bibfield  {journal} {\bibinfo  {journal} {Phys. Rev. Lett.}\ }\textbf
  {\bibinfo {volume} {98}},\ \bibinfo {pages} {131104} (\bibinfo {year}
  {2007})}\BibitemShut {NoStop}%
\bibitem [{\citenamefont {Chen}\ \emph {et~al.}(2016)\citenamefont {Chen},
  \citenamefont {Tham}, \citenamefont {Krause}, \citenamefont {Lopez},
  \citenamefont {Fischbach},\ and\ \citenamefont {Decca}}]{Chen2016}%
  \BibitemOpen
  \bibfield  {author} {\bibinfo {author} {\bibfnamefont {Y.-J.}\ \bibnamefont
  {Chen}}, \bibinfo {author} {\bibfnamefont {W.~K.}\ \bibnamefont {Tham}},
  \bibinfo {author} {\bibfnamefont {D.~E.}\ \bibnamefont {Krause}}, \bibinfo
  {author} {\bibfnamefont {D.}~\bibnamefont {Lopez}}, \bibinfo {author}
  {\bibfnamefont {E.}~\bibnamefont {Fischbach}},\ and\ \bibinfo {author}
  {\bibfnamefont {R.~S.}\ \bibnamefont {Decca}},\ }\href@noop {} {\bibfield
  {journal} {\bibinfo  {journal} {Phys. Rev. Lett.}\ }\textbf {\bibinfo
  {volume} {116}},\ \bibinfo {pages} {221102} (\bibinfo {year}
  {2016})}\BibitemShut {NoStop}%
\bibitem [{\citenamefont {Vasilakis}\ \emph {et~al.}(2009)\citenamefont
  {Vasilakis}, \citenamefont {Brown}, \citenamefont {Kornack},\ and\
  \citenamefont {Romalis}}]{Vasilakis2009}%
  \BibitemOpen
  \bibfield  {author} {\bibinfo {author} {\bibfnamefont {G.}~\bibnamefont
  {Vasilakis}}, \bibinfo {author} {\bibfnamefont {J.~M.}\ \bibnamefont
  {Brown}}, \bibinfo {author} {\bibfnamefont {T.}~\bibnamefont {Kornack}},\
  and\ \bibinfo {author} {\bibfnamefont {M.~V.}\ \bibnamefont {Romalis}},\
  }\href@noop {} {\bibfield  {journal} {\bibinfo  {journal} {Phys. Rev. Lett.}\
  }\textbf {\bibinfo {volume} {103}},\ \bibinfo {pages} {261801} (\bibinfo
  {year} {2009})}\BibitemShut {NoStop}%
\bibitem [{\citenamefont {Terrano}\ \emph {et~al.}(2015)\citenamefont
  {Terrano}, \citenamefont {Adelberger}, \citenamefont {Lee},\ and\
  \citenamefont {Heckel}}]{Terrano2015}%
  \BibitemOpen
  \bibfield  {author} {\bibinfo {author} {\bibfnamefont {W.~A.}\ \bibnamefont
  {Terrano}}, \bibinfo {author} {\bibfnamefont {E.~G.}\ \bibnamefont
  {Adelberger}}, \bibinfo {author} {\bibfnamefont {J.~G.}\ \bibnamefont
  {Lee}},\ and\ \bibinfo {author} {\bibfnamefont {B.~R.}\ \bibnamefont
  {Heckel}},\ }\href@noop {} {\bibfield  {journal} {\bibinfo  {journal} {Phys.
  Rev. Lett.}\ }\textbf {\bibinfo {volume} {115}},\ \bibinfo {pages} {201801}
  (\bibinfo {year} {2015})}\BibitemShut {NoStop}%
\bibitem [{\citenamefont {Feng}\ and\ \citenamefont
  {Hong-Ya}(2006)}]{FengCPL2006}%
  \BibitemOpen
  \bibfield  {author} {\bibinfo {author} {\bibfnamefont {L.}~\bibnamefont
  {Feng}}\ and\ \bibinfo {author} {\bibfnamefont {L.}~\bibnamefont {Hong-Ya}},\
  }\href {https://doi.org/10.1088/0256-307x/23/11/006} {\bibfield  {journal}
  {\bibinfo  {journal} {Chinese Physics Letters}\ }\textbf {\bibinfo {volume}
  {23}},\ \bibinfo {pages} {2903} (\bibinfo {year} {2006})}\BibitemShut
  {NoStop}%
\bibitem [{\citenamefont {Luo}\ and\ \citenamefont {Liu}(2007)}]{FengIJTP2007}%
  \BibitemOpen
  \bibfield  {author} {\bibinfo {author} {\bibfnamefont {F.}~\bibnamefont
  {Luo}}\ and\ \bibinfo {author} {\bibfnamefont {H.}~\bibnamefont {Liu}},\
  }\href {https://doi.org/10.1007/s10773-006-9164-6} {\bibfield  {journal}
  {\bibinfo  {journal} {Int. J. Theor. Phys.}\ }\textbf {\bibinfo {volume}
  {46}},\ \bibinfo {pages} {606} (\bibinfo {year} {2007})},\ \Eprint
  {https://arxiv.org/abs/gr-qc/0602093} {arXiv:gr-qc/0602093} \BibitemShut
  {NoStop}%
\bibitem [{\citenamefont {Dahia}\ and\ \citenamefont
  {Lemos}(2016)}]{DahiaPRD2016}%
  \BibitemOpen
  \bibfield  {author} {\bibinfo {author} {\bibfnamefont {F.}~\bibnamefont
  {Dahia}}\ and\ \bibinfo {author} {\bibfnamefont {A.~S.}\ \bibnamefont
  {Lemos}},\ }\href {https://doi.org/10.1103/PhysRevD.94.084033} {\bibfield
  {journal} {\bibinfo  {journal} {Phys. Rev. D}\ }\textbf {\bibinfo {volume}
  {94}},\ \bibinfo {pages} {084033} (\bibinfo {year} {2016})}\BibitemShut
  {NoStop}%
\bibitem [{\citenamefont {Zhi-Gang}\ \emph {et~al.}(2008)\citenamefont
  {Zhi-Gang}, \citenamefont {Wei-Tou},\ and\ \citenamefont
  {Pat{\'{o}}n}}]{ZhiGang2008}%
  \BibitemOpen
  \bibfield  {author} {\bibinfo {author} {\bibfnamefont {L.}~\bibnamefont
  {Zhi-Gang}}, \bibinfo {author} {\bibfnamefont {N.}~\bibnamefont {Wei-Tou}},\
  and\ \bibinfo {author} {\bibfnamefont {A.~P.}\ \bibnamefont {Pat{\'{o}}n}},\
  }\href {https://doi.org/10.1088/1674-1056/17/1/013} {\bibfield  {journal}
  {\bibinfo  {journal} {Chinese Physics B}\ }\textbf {\bibinfo {volume} {17}},\
  \bibinfo {pages} {70} (\bibinfo {year} {2008})}\BibitemShut {NoStop}%
\bibitem [{\citenamefont {Wang}\ and\ \citenamefont {Ni}(2013)}]{WangMPLA2013}%
  \BibitemOpen
  \bibfield  {author} {\bibinfo {author} {\bibfnamefont {L.-B.}\ \bibnamefont
  {Wang}}\ and\ \bibinfo {author} {\bibfnamefont {W.-T.}\ \bibnamefont {Ni}},\
  }\href {https://doi.org/10.1142/s0217732313500946} {\bibfield  {journal}
  {\bibinfo  {journal} {Modern Physics Letters A}\ }\textbf {\bibinfo {volume}
  {28}},\ \bibinfo {pages} {1350094} (\bibinfo {year} {2013})}\BibitemShut
  {NoStop}%
\bibitem [{\citenamefont {Dzuba}\ \emph {et~al.}(2022)\citenamefont {Dzuba},
  \citenamefont {Flambaum},\ and\ \citenamefont {Munro-Laylim}}]{MunroLaylim}%
  \BibitemOpen
  \bibfield  {author} {\bibinfo {author} {\bibfnamefont {V.~A.}\ \bibnamefont
  {Dzuba}}, \bibinfo {author} {\bibfnamefont {V.~V.}\ \bibnamefont
  {Flambaum}},\ and\ \bibinfo {author} {\bibfnamefont {P.}~\bibnamefont
  {Munro-Laylim}},\ }\href {https://doi.org/10.1103/PhysRevA.106.052804}
  {\bibfield  {journal} {\bibinfo  {journal} {Phys. Rev. A}\ }\textbf {\bibinfo
  {volume} {106}},\ \bibinfo {pages} {052804} (\bibinfo {year}
  {2022})}\BibitemShut {NoStop}%
\bibitem [{\citenamefont {Salumbides}\ \emph {et~al.}(2015)\citenamefont
  {Salumbides}, \citenamefont {Schellekens}, \citenamefont {Gato-Rivera},\ and\
  \citenamefont {Ubachs}}]{Salumbides2015}%
  \BibitemOpen
  \bibfield  {author} {\bibinfo {author} {\bibfnamefont {E.~J.}\ \bibnamefont
  {Salumbides}}, \bibinfo {author} {\bibfnamefont {A.~N.}\ \bibnamefont
  {Schellekens}}, \bibinfo {author} {\bibfnamefont {B.}~\bibnamefont
  {Gato-Rivera}},\ and\ \bibinfo {author} {\bibfnamefont {W.}~\bibnamefont
  {Ubachs}},\ }\href {https://doi.org/10.1088/1367-2630/17/3/033015} {\bibfield
   {journal} {\bibinfo  {journal} {New Journal of Physics}\ }\textbf {\bibinfo
  {volume} {17}},\ \bibinfo {pages} {033015} (\bibinfo {year}
  {2015})}\BibitemShut {NoStop}%
\bibitem [{\citenamefont {Witten}(1984)}]{Witten84}%
  \BibitemOpen
  \bibfield  {author} {\bibinfo {author} {\bibfnamefont {E.}~\bibnamefont
  {Witten}},\ }\href {https://doi.org/10.1103/PhysRevD.30.272} {\bibfield
  {journal} {\bibinfo  {journal} {Phys. Rev. D}\ }\textbf {\bibinfo {volume}
  {30}},\ \bibinfo {pages} {272} (\bibinfo {year} {1984})}\BibitemShut
  {NoStop}%
\bibitem [{\citenamefont {Farhi}\ and\ \citenamefont {Jaffe}(1984)}]{Farhi}%
  \BibitemOpen
  \bibfield  {author} {\bibinfo {author} {\bibfnamefont {E.}~\bibnamefont
  {Farhi}}\ and\ \bibinfo {author} {\bibfnamefont {R.~L.}\ \bibnamefont
  {Jaffe}},\ }\href {https://doi.org/10.1103/PhysRevD.30.2379} {\bibfield
  {journal} {\bibinfo  {journal} {Phys. Rev. D}\ }\textbf {\bibinfo {volume}
  {30}},\ \bibinfo {pages} {2379} (\bibinfo {year} {1984})}\BibitemShut
  {NoStop}%
\bibitem [{\citenamefont {De~Rujula}\ and\ \citenamefont
  {Glashow}(1984)}]{Glashow}%
  \BibitemOpen
  \bibfield  {author} {\bibinfo {author} {\bibfnamefont {A.}~\bibnamefont
  {De~Rujula}}\ and\ \bibinfo {author} {\bibfnamefont {S.~L.}\ \bibnamefont
  {Glashow}},\ }\href {https://doi.org/10.1038/312734a0} {\bibfield  {journal}
  {\bibinfo  {journal} {Nature}\ }\textbf {\bibinfo {volume} {312}},\ \bibinfo
  {pages} {734} (\bibinfo {year} {1984})}\BibitemShut {NoStop}%
\bibitem [{\citenamefont {Jacobs}\ \emph
  {et~al.}(2015{\natexlab{a}})\citenamefont {Jacobs}, \citenamefont
  {Starkman},\ and\ \citenamefont {Lynn}}]{Starkman1}%
  \BibitemOpen
  \bibfield  {author} {\bibinfo {author} {\bibfnamefont {D.~M.}\ \bibnamefont
  {Jacobs}}, \bibinfo {author} {\bibfnamefont {G.~D.}\ \bibnamefont
  {Starkman}},\ and\ \bibinfo {author} {\bibfnamefont {B.~W.}\ \bibnamefont
  {Lynn}},\ }\href {https://doi.org/10.1093/mnras/stv774} {\bibfield  {journal}
  {\bibinfo  {journal} {Month. Not. Royal Astronom. Soc.}\ }\textbf {\bibinfo
  {volume} {450}},\ \bibinfo {pages} {3418} (\bibinfo {year}
  {2015}{\natexlab{a}})}\BibitemShut {NoStop}%
\bibitem [{\citenamefont {Jacobs}\ \emph
  {et~al.}(2015{\natexlab{b}})\citenamefont {Jacobs}, \citenamefont {Weltman},\
  and\ \citenamefont {Starkman}}]{Starkman2}%
  \BibitemOpen
  \bibfield  {author} {\bibinfo {author} {\bibfnamefont {D.~M.}\ \bibnamefont
  {Jacobs}}, \bibinfo {author} {\bibfnamefont {A.}~\bibnamefont {Weltman}},\
  and\ \bibinfo {author} {\bibfnamefont {G.~D.}\ \bibnamefont {Starkman}},\
  }\href {https://doi.org/10.1103/PhysRevD.91.115023} {\bibfield  {journal}
  {\bibinfo  {journal} {Phys. Rev. D}\ }\textbf {\bibinfo {volume} {91}},\
  \bibinfo {pages} {115023} (\bibinfo {year} {2015}{\natexlab{b}})}\BibitemShut
  {NoStop}%
\bibitem [{\citenamefont {Sidhu}\ \emph
  {et~al.}(2019{\natexlab{a}})\citenamefont {Sidhu}, \citenamefont {Abraham},
  \citenamefont {Covault},\ and\ \citenamefont {Starkman}}]{Starkman3}%
  \BibitemOpen
  \bibfield  {author} {\bibinfo {author} {\bibfnamefont {J.~S.}\ \bibnamefont
  {Sidhu}}, \bibinfo {author} {\bibfnamefont {R.~M.}\ \bibnamefont {Abraham}},
  \bibinfo {author} {\bibfnamefont {C.}~\bibnamefont {Covault}},\ and\ \bibinfo
  {author} {\bibfnamefont {G.}~\bibnamefont {Starkman}},\ }\href
  {https://doi.org/10.1088/1475-7516/2019/02/037} {\bibfield  {journal}
  {\bibinfo  {journal} {JCAP}\ }\textbf {\bibinfo {volume} {2019}}\bibinfo
  {number} { (02)},\ \bibinfo {pages} {037}}\BibitemShut {NoStop}%
\bibitem [{\citenamefont {Sidhu}\ \emph
  {et~al.}(2019{\natexlab{b}})\citenamefont {Sidhu}, \citenamefont {Starkman},\
  and\ \citenamefont {Harvey}}]{Starkman4}%
  \BibitemOpen
\bibfield  {number} {  }\bibfield  {author} {\bibinfo {author} {\bibfnamefont
  {J.~S.}\ \bibnamefont {Sidhu}}, \bibinfo {author} {\bibfnamefont
  {G.}~\bibnamefont {Starkman}},\ and\ \bibinfo {author} {\bibfnamefont
  {R.}~\bibnamefont {Harvey}},\ }\href
  {https://doi.org/10.1103/PhysRevD.100.103015} {\bibfield  {journal} {\bibinfo
   {journal} {Phys. Rev. D}\ }\textbf {\bibinfo {volume} {100}},\ \bibinfo
  {pages} {103015} (\bibinfo {year} {2019}{\natexlab{b}})}\BibitemShut
  {NoStop}%
\bibitem [{\citenamefont {Sidhu}\ \emph {et~al.}(2020)\citenamefont {Sidhu},
  \citenamefont {Scherrer},\ and\ \citenamefont {Starkman}}]{Starkman5}%
  \BibitemOpen
  \bibfield  {author} {\bibinfo {author} {\bibfnamefont {J.~S.}\ \bibnamefont
  {Sidhu}}, \bibinfo {author} {\bibfnamefont {R.}~\bibnamefont {Scherrer}},\
  and\ \bibinfo {author} {\bibfnamefont {G.}~\bibnamefont {Starkman}},\ }\href
  {https://doi.org/https://doi.org/10.1016/j.physletb.2020.135300} {\bibfield
  {journal} {\bibinfo  {journal} {Phys. Lett. B}\ }\textbf {\bibinfo {volume}
  {803}},\ \bibinfo {pages} {135300} (\bibinfo {year} {2020})}\BibitemShut
  {NoStop}%
\bibitem [{\citenamefont {Sidhu}\ and\ \citenamefont
  {Starkman}(2019)}]{Starkman6}%
  \BibitemOpen
  \bibfield  {author} {\bibinfo {author} {\bibfnamefont {J.~S.}\ \bibnamefont
  {Sidhu}}\ and\ \bibinfo {author} {\bibfnamefont {G.}~\bibnamefont
  {Starkman}},\ }\href {https://doi.org/10.1103/PhysRevD.100.123008} {\bibfield
   {journal} {\bibinfo  {journal} {Phys. Rev. D}\ }\textbf {\bibinfo {volume}
  {100}},\ \bibinfo {pages} {123008} (\bibinfo {year} {2019})}\BibitemShut
  {NoStop}%
\bibitem [{\citenamefont {Sidhu}\ and\ \citenamefont
  {Starkman}(2020)}]{Starkman7}%
  \BibitemOpen
  \bibfield  {author} {\bibinfo {author} {\bibfnamefont {J.~S.}\ \bibnamefont
  {Sidhu}}\ and\ \bibinfo {author} {\bibfnamefont {G.~D.}\ \bibnamefont
  {Starkman}},\ }\href {https://doi.org/10.1103/PhysRevD.101.083503} {\bibfield
   {journal} {\bibinfo  {journal} {Phys. Rev. D}\ }\textbf {\bibinfo {volume}
  {101}},\ \bibinfo {pages} {083503} (\bibinfo {year} {2020})}\BibitemShut
  {NoStop}%
\bibitem [{\citenamefont {Sidhu}(2020)}]{Starkman8}%
  \BibitemOpen
  \bibfield  {author} {\bibinfo {author} {\bibfnamefont {J.~S.}\ \bibnamefont
  {Sidhu}},\ }\href {https://doi.org/10.1103/PhysRevD.101.043526} {\bibfield
  {journal} {\bibinfo  {journal} {Phys. Rev. D}\ }\textbf {\bibinfo {volume}
  {101}},\ \bibinfo {pages} {043526} (\bibinfo {year} {2020})}\BibitemShut
  {NoStop}%
\bibitem [{\citenamefont {Cyncynates}\ \emph {et~al.}(2017)\citenamefont
  {Cyncynates}, \citenamefont {Chiel}, \citenamefont {Sidhu},\ and\
  \citenamefont {Starkman}}]{Starkman9}%
  \BibitemOpen
  \bibfield  {author} {\bibinfo {author} {\bibfnamefont {D.}~\bibnamefont
  {Cyncynates}}, \bibinfo {author} {\bibfnamefont {J.}~\bibnamefont {Chiel}},
  \bibinfo {author} {\bibfnamefont {J.}~\bibnamefont {Sidhu}},\ and\ \bibinfo
  {author} {\bibfnamefont {G.~D.}\ \bibnamefont {Starkman}},\ }\href
  {https://doi.org/10.1103/PhysRevD.95.063006} {\bibfield  {journal} {\bibinfo
  {journal} {Phys. Rev. D}\ }\textbf {\bibinfo {volume} {95}},\ \bibinfo
  {pages} {063006} (\bibinfo {year} {2017})}\BibitemShut {NoStop}%
\bibitem [{\citenamefont {Singh~Sidhu}\ \emph {et~al.}(2020)\citenamefont
  {Singh~Sidhu}, \citenamefont {Scherrer},\ and\ \citenamefont
  {Starkman}}]{Starkman10}%
  \BibitemOpen
  \bibfield  {author} {\bibinfo {author} {\bibfnamefont {J.}~\bibnamefont
  {Singh~Sidhu}}, \bibinfo {author} {\bibfnamefont {R.~J.}\ \bibnamefont
  {Scherrer}},\ and\ \bibinfo {author} {\bibfnamefont {G.}~\bibnamefont
  {Starkman}},\ }\href {https://doi.org/10.1016/j.physletb.2020.135574}
  {\bibfield  {journal} {\bibinfo  {journal} {Phys. Lett. B}\ }\textbf
  {\bibinfo {volume} {807}},\ \bibinfo {pages} {135574} (\bibinfo {year}
  {2020})}\BibitemShut {NoStop}%
\bibitem [{\citenamefont {Caloni}\ \emph {et~al.}(2021)\citenamefont {Caloni},
  \citenamefont {Gerbino},\ and\ \citenamefont {Lattanzi}}]{Caloni}%
  \BibitemOpen
  \bibfield  {author} {\bibinfo {author} {\bibfnamefont {L.}~\bibnamefont
  {Caloni}}, \bibinfo {author} {\bibfnamefont {M.}~\bibnamefont {Gerbino}},\
  and\ \bibinfo {author} {\bibfnamefont {M.}~\bibnamefont {Lattanzi}},\ }\href
  {https://doi.org/10.1088/1475-7516/2021/07/027} {\bibfield  {journal}
  {\bibinfo  {journal} {JCAP}\ }\textbf {\bibinfo {volume} {07}},\ \bibinfo
  {pages} {027}}\BibitemShut {NoStop}%
\bibitem [{\citenamefont {Flambaum}\ and\ \citenamefont
  {Samsonov}(2021)}]{FS0}%
  \BibitemOpen
  \bibfield  {author} {\bibinfo {author} {\bibfnamefont {V.~V.}\ \bibnamefont
  {Flambaum}}\ and\ \bibinfo {author} {\bibfnamefont {I.~B.}\ \bibnamefont
  {Samsonov}},\ }\href {https://doi.org/10.1103/PhysRevD.104.063042} {\bibfield
   {journal} {\bibinfo  {journal} {Phys. Rev. D}\ }\textbf {\bibinfo {volume}
  {104}},\ \bibinfo {pages} {063042} (\bibinfo {year} {2021})}\BibitemShut
  {NoStop}%
\bibitem [{\citenamefont {Flambaum}\ and\ \citenamefont
  {Samsonov}(2022{\natexlab{a}})}]{FS1}%
  \BibitemOpen
  \bibfield  {author} {\bibinfo {author} {\bibfnamefont {V.~V.}\ \bibnamefont
  {Flambaum}}\ and\ \bibinfo {author} {\bibfnamefont {I.~B.}\ \bibnamefont
  {Samsonov}},\ }\href {https://doi.org/10.1103/PhysRevD.105.123011} {\bibfield
   {journal} {\bibinfo  {journal} {Phys. Rev. D}\ }\textbf {\bibinfo {volume}
  {105}},\ \bibinfo {pages} {123011} (\bibinfo {year}
  {2022}{\natexlab{a}})}\BibitemShut {NoStop}%
\bibitem [{\citenamefont {Flambaum}\ and\ \citenamefont
  {Samsonov}(2022{\natexlab{b}})}]{FS2}%
  \BibitemOpen
  \bibfield  {author} {\bibinfo {author} {\bibfnamefont {V.~V.}\ \bibnamefont
  {Flambaum}}\ and\ \bibinfo {author} {\bibfnamefont {I.~B.}\ \bibnamefont
  {Samsonov}},\ }\href {https://doi.org/10.1103/PhysRevD.106.023006} {\bibfield
   {journal} {\bibinfo  {journal} {Phys. Rev. D}\ }\textbf {\bibinfo {volume}
  {106}},\ \bibinfo {pages} {023006} (\bibinfo {year}
  {2022}{\natexlab{b}})}\BibitemShut {NoStop}%
  
\bibitem [{\citenamefont {Flambaum}\ \emph {et~al.}(2023)\citenamefont
  {Flambaum}, \citenamefont {Samsonov},\ and\ \citenamefont {Vong}}]{Garry}%
  \BibitemOpen
  \bibfield  {author} {\bibinfo {author} {\bibfnamefont {V.~V.}\ \bibnamefont
  {Flambaum}}, \bibinfo {author} {\bibfnamefont {I.~B.}\ \bibnamefont
  {Samsonov}},\ and\ \bibinfo {author} {\bibfnamefont {G.~K.}\ \bibnamefont
  {Vong}},\ }\href {https://doi.org/10.1103/PhysRevD.107.123501} {\bibfield
  {journal} {\bibinfo  {journal} {Phys. Rev. D}\ }\textbf {\bibinfo {volume}
  {107}},\ \bibinfo {pages} {123501} (\bibinfo {year} {2023})}\BibitemShut
  {NoStop}%

\bibitem{Garry2025} 
V. V. Flambaum, I. B. Samsonov, G. K. Vong, Phys. Rev. D {\bf 111}, 023525, (2025). DOI: 10.1103/PhysRevD.111.023525
 
  \bibitem{Budker2020}
   Dmitry Budker, Victor V.
Flambaum, Xunyu Liang, Ariel Zhitnitsky, Phys. Rev. D {\bf 101}, 043012 (2020).
  DOI: 10.1103/PhysRevD.101.043012, arxiv:1909.09475

\bibitem [{\citenamefont {Budker}\ \emph {et~al.}(2022)\citenamefont {Budker},
  \citenamefont {Flambaum},\ and\ \citenamefont {Zhitnitsky}}]{Budker2022}%
  \BibitemOpen
  \bibfield  {author} {\bibinfo {author} {\bibfnamefont {D.}~\bibnamefont
  {Budker}}, \bibinfo {author} {\bibfnamefont {V.~V.}\ \bibnamefont
  {Flambaum}},\ and\ \bibinfo {author} {\bibfnamefont {A.}~\bibnamefont
  {Zhitnitsky}},\ }\href {https://doi.org/10.3390/sym14030459} {\bibfield
  {journal} {\bibinfo  {journal} {Symmetry}\ }\textbf {\bibinfo {volume}
  {14}},\ \bibinfo {pages} {459} (\bibinfo {year} {2022})}\BibitemShut
  {NoStop}%
\bibitem [{\citenamefont {Forbes}\ and\ \citenamefont
  {Zhitnitsky}(2001)}]{DW1}%
  \BibitemOpen
  \bibfield  {author} {\bibinfo {author} {\bibfnamefont {M.~M.}\ \bibnamefont
  {Forbes}}\ and\ \bibinfo {author} {\bibfnamefont {A.~R.}\ \bibnamefont
  {Zhitnitsky}},\ }\href {https://doi.org/10.1088/1126-6708/2001/10/013}
  {\bibfield  {journal} {\bibinfo  {journal} {JHEP}\ }\textbf {\bibinfo
  {volume} {2001}}\bibinfo  {number} { (10)},\ \bibinfo {pages}
  {013}}\BibitemShut {NoStop}%
\bibitem [{\citenamefont {Son}\ \emph {et~al.}(2001)\citenamefont {Son},
  \citenamefont {Stephanov},\ and\ \citenamefont {Zhitnitsky}}]{DW2}%
  \BibitemOpen
\bibfield  {number} {  }\bibfield  {author} {\bibinfo {author} {\bibfnamefont
  {D.~T.}\ \bibnamefont {Son}}, \bibinfo {author} {\bibfnamefont {M.~A.}\
  \bibnamefont {Stephanov}},\ and\ \bibinfo {author} {\bibfnamefont {A.~R.}\
  \bibnamefont {Zhitnitsky}},\ }\href
  {https://doi.org/10.1103/PhysRevLett.86.3955} {\bibfield  {journal} {\bibinfo
   {journal} {Phys. Rev. Lett.}\ }\textbf {\bibinfo {volume} {86}},\ \bibinfo
  {pages} {3955} (\bibinfo {year} {2001})}\BibitemShut {NoStop}%
\bibitem [{\citenamefont {Shuryak}\ and\ \citenamefont
  {Zhitnitsky}(2002)}]{DW3}%
  \BibitemOpen
  \bibfield  {author} {\bibinfo {author} {\bibfnamefont {E.~V.}\ \bibnamefont
  {Shuryak}}\ and\ \bibinfo {author} {\bibfnamefont {A.~R.}\ \bibnamefont
  {Zhitnitsky}},\ }\href {https://doi.org/10.1103/PhysRevC.66.034905}
  {\bibfield  {journal} {\bibinfo  {journal} {Phys. Rev. C}\ }\textbf {\bibinfo
  {volume} {66}},\ \bibinfo {pages} {034905} (\bibinfo {year}
  {2002})}\BibitemShut {NoStop}%
\bibitem [{\citenamefont {Zhitnitsky}(2003)}]{Zhitnitsky_2003}%
  \BibitemOpen
  \bibfield  {author} {\bibinfo {author} {\bibfnamefont {A.~R.}\ \bibnamefont
  {Zhitnitsky}},\ }\href {https://doi.org/10.1088/1475-7516/2003/10/010}
  {\bibfield  {journal} {\bibinfo  {journal} {JCAP}\ }\textbf {\bibinfo
  {volume} {2003}}\bibinfo  {number} { (10)},\ \bibinfo {pages}
  {010}}\BibitemShut {NoStop}%
\bibitem [{\citenamefont {Zhitnitsky}(2006)}]{CCO}%
  \BibitemOpen
\bibfield  {number} {  }\bibfield  {author} {\bibinfo {author} {\bibfnamefont
  {A.}~\bibnamefont {Zhitnitsky}},\ }\href
  {https://doi.org/10.1103/PhysRevD.74.043515} {\bibfield  {journal} {\bibinfo
  {journal} {Phys. Rev. D}\ }\textbf {\bibinfo {volume} {74}},\ \bibinfo
  {pages} {043515} (\bibinfo {year} {2006})}\BibitemShut {NoStop}%
\bibitem [{\citenamefont {Oaknin}\ and\ \citenamefont
  {Zhitnitsky}(2005)}]{Basymmetry}%
  \BibitemOpen
  \bibfield  {author} {\bibinfo {author} {\bibfnamefont {D.~H.}\ \bibnamefont
  {Oaknin}}\ and\ \bibinfo {author} {\bibfnamefont {A.}~\bibnamefont
  {Zhitnitsky}},\ }\href {https://doi.org/10.1103/PhysRevD.71.023519}
  {\bibfield  {journal} {\bibinfo  {journal} {Phys. Rev. D}\ }\textbf {\bibinfo
  {volume} {71}},\ \bibinfo {pages} {023519} (\bibinfo {year}
  {2005})}\BibitemShut {NoStop}%
\bibitem [{\citenamefont {Forbes}\ and\ \citenamefont
  {Zhitnitsky}(2008)}]{WMAPhaze}%
  \BibitemOpen
  \bibfield  {author} {\bibinfo {author} {\bibfnamefont {M.~M.}\ \bibnamefont
  {Forbes}}\ and\ \bibinfo {author} {\bibfnamefont {A.~R.}\ \bibnamefont
  {Zhitnitsky}},\ }\href {https://doi.org/10.1103/PhysRevD.78.083505}
  {\bibfield  {journal} {\bibinfo  {journal} {Phys. Rev. D}\ }\textbf {\bibinfo
  {volume} {78}},\ \bibinfo {pages} {083505} (\bibinfo {year}
  {2008})}\BibitemShut {NoStop}%
\bibitem [{\citenamefont {Ruffini}\ and\ \citenamefont
  {Bonazzola}(1969)}]{Ruffini}%
  \BibitemOpen
  \bibfield  {author} {\bibinfo {author} {\bibfnamefont {R.}~\bibnamefont
  {Ruffini}}\ and\ \bibinfo {author} {\bibfnamefont {S.}~\bibnamefont
  {Bonazzola}},\ }\href {https://doi.org/10.1103/PhysRev.187.1767} {\bibfield
  {journal} {\bibinfo  {journal} {Phys. Rev.}\ }\textbf {\bibinfo {volume}
  {187}},\ \bibinfo {pages} {1767} (\bibinfo {year} {1969})}\BibitemShut
  {NoStop}%
\bibitem [{\citenamefont {Flambaum}\ and\ \citenamefont
  {Samsonov}(2024)}]{BoseStar}%
  \BibitemOpen
  \bibfield  {author} {\bibinfo {author} {\bibfnamefont {V.~V.}\ \bibnamefont
  {Flambaum}}\ and\ \bibinfo {author} {\bibfnamefont {I.~B.}\ \bibnamefont
  {Samsonov}},\ }\href {https://doi.org/10.1103/PhysRevD.110.103016} {\bibfield
   {journal} {\bibinfo  {journal} {Phys. Rev. D}\ }\textbf {\bibinfo {volume}
  {110}},\ \bibinfo {pages} {103016} (\bibinfo {year} {2024})},\ \Eprint
  {https://arxiv.org/abs/2407.21262} {arXiv:2407.21262 [hep-ph]} \BibitemShut
  {NoStop}%
\bibitem [{\citenamefont {Landau}\ and\ \citenamefont
  {Lifshitz}(1980)}]{Landau5}%
  \BibitemOpen
  \bibfield  {author} {\bibinfo {author} {\bibfnamefont {L.~D.}\ \bibnamefont
  {Landau}}\ and\ \bibinfo {author} {\bibfnamefont {E.~M.}\ \bibnamefont
  {Lifshitz}},\ }\href@noop {} {\emph {\bibinfo {title} {Statistical Physics,
  part 1}}}\ (\bibinfo  {publisher} {{Pergamon press, Oxford}},\ \bibinfo
  {year} {1980})\BibitemShut {NoStop}%
\bibitem [{\citenamefont {Workman}\ and\ \citenamefont {Others}(2022)}]{PDG}%
  \BibitemOpen
  \bibfield  {author} {\bibinfo {author} {\bibfnamefont {R.~L.}\ \bibnamefont
  {Workman}}\ and\ \bibinfo {author} {\bibnamefont {Others}} (\bibinfo
  {collaboration} {Particle Data Group}),\ }\href
  {https://doi.org/10.1093/ptep/ptac097} {\bibfield  {journal} {\bibinfo
  {journal} {PTEP}\ }\textbf {\bibinfo {volume} {2022}},\ \bibinfo {pages}
  {083C01} (\bibinfo {year} {2022})}\BibitemShut {NoStop}%
\bibitem [{\citenamefont {Mark~Vogelsberger}\ and\ \citenamefont
  {Loeb}(2012)}]{DMselfinteraction}%
  \BibitemOpen
  \bibfield  {author} {\bibinfo {author} {\bibfnamefont {J.~Z.}\ \bibnamefont
  {Mark~Vogelsberger}}\ and\ \bibinfo {author} {\bibfnamefont {A.}~\bibnamefont
  {Loeb}},\ }\href {https://doi.org/10.1111/j.1365-2966.2012.21182.x}
  {\bibfield  {journal} {\bibinfo  {journal} {Monthly Notices of the Royal
  Astronomical Society}\ }\textbf {\bibinfo {volume} {423}},\ \bibinfo {pages}
  {3740} (\bibinfo {year} {2012})}\BibitemShut {NoStop}%
  
  
\bibitem{BlackHole}   I. Musco and K. Jedamzik,
   Phys. Rev. D {bf 103,} 063538 (2021).
   
\bibitem{neutronevaporation}    C. Alcock and E. Farhi, Phys. Rev. D {\bf 32}, 1273 (1985).
   
   \bibitem{MadsenHeiselberg}J. Madsen, H. Heiselberg and K. Riisager,  Phys. Rev. D {\bf 34},  2947 (1986).
   
  \bibitem{Madsen1998}  J. Madsen, Physics and astrophysics of strange quark matter, 1998 Preprint astro-ph/9809032
  
   
\bibitem{Zhitnitsky2019}   S. Ge, K. Lawson, and A. Zhitnitsky, Phys. Rev. D {\bf 99}, 116017 (2019).

\bibitem{ZhitnitskyReview} A. Zhitnitsky, Mod. Phys. Lett. A {\bf 36}, 2130017 (2021).

\bibitem{Radar} P. Dhakal, S. Prohira, C. V. Cappiello, J. F. Beacom, S. Palo,
and J. Marino, Phys. Rev. D {\bf 107}, 043026 (2023).
\bibitem{Herrin1} E. T. Herrin, D. C. Rosenbaum, and V. L. Teplitz, Phys. Rev.
D {\bf 73}, 043511 (2006).
\bibitem{Herrin2} E E. T. Herrin and V. L. Teplitz, Phys. Rev. D {\bf 53}, 6762 (1996).
\bibitem{Herrin3} E D. P. Anderson, E. T. Herrin, V. L. Teplitz, and I. M.
Tibuleac, Bull. Seismol. Soc. Am. {\bf 93}, 2363 (2003).

\end{thebibliography}
\end{document}